\documentclass[12pt]{article}

\usepackage{epsfig}
\usepackage{graphicx}

\usepackage{graphicx}    

\def\beq{\begin{equation}}
\def\eeq{\end{equation}}

\def\bea{\begin{eqnarray}}
\def\eea{\end{eqnarray}}
\def\ba{\begin{array}}                  
\def\ea{\end{array}}

\begin{document}

\begin{titlepage}

\rightline{UB-ECM-PF-05/02}

\rightline{hep-th/0501132}

\vskip 2cm
\begin{center}

{\Large \bf The Information Problem in Black Hole Evaporation: Old and
  Recent Results\footnote{ Based
on Talks given at  ERE2004 ``Beyond General Relativity'', Miraflores
de la Sierra, Madrid (Sept 2004), and at CERN (Oct 2004).}}

\vskip  2 cm

{\large Jorge G. Russo  }

\end{center}

\bigskip

\centerline {Instituci\'o Catalana de Recerca i Estudis
Avan\c cats (ICREA),}

\centerline{ Departament ECM, Facultat de F\'{\i}sica, Universitat de
  Barcelona,  Spain.}

%
%
%

\bigskip

\begin{abstract}

The formation and evaporation of a black hole can be viewed as a
scattering process in
Quantum Gravity. Semiclassical arguments indicate that the process should
be non-unitary,
and that all the information of the original quantum state forming the
black hole should be lost after  the black hole has completely evaporated, except
for its mass, charge and angular momentum.
This would imply  a violation of basic principles of quantum mechanics.
 We review some proposed resolutions to the problem,
including developments in string theory and a recent proposal by Hawking.
We also suggest a novel approach which makes use of some ingredients 
of earlier proposals.

\end{abstract}
\bigskip
\bigskip
\bigskip
\bigskip\bigskip

\noindent {January 2005}

\vfill

\end{titlepage}

\setcounter{section}{0}
\section{Introduction}
\label{sec:1}

\setcounter{equation}{0}

Black holes emit thermal radiation  
with a spectrum given by \cite{Hawk75}
\begin{equation}
\Gamma (\omega )= {\sigma (\omega )\over 
e^{\omega\over T_H}-1}\omega^{d-2} d\omega \ ,
\end{equation}
where $T_H$ is the Hawking temperature which depends on the mass, charge and angular momentum, $d$ is the number of spatial dimensions and $\sigma (\omega )$
is a grey body factor.
The process continues until the black hole  completely evaporates, leaving behind only thermal radiation. 
All the information about the original quantum state that formed the black hole seems to be lost, except
for its mass, charge and angular momentum.


Black hole formation and evaporation can be thought of as scattering process.
According to Hawking \cite{Hawk76}, all correlators with the infalling matter that formed the black hole decay exponentially with time. 
Pure states seem to evolve into mixed states, contrary to the basic principles of quantum mechanics.

At Planck energies, black holes are expected to appear virtually in
many scattering processes. This would require changes in 
the rules of quantum mechanics. In particular, if pure states evolve
into mixed states, the S matrix would not exist.
Because the final state is described by a density matrix, 
in  \cite{Hawk76} Hawking proposed  to introduce a ``superscattering'' operator 
that maps density matrices to density matrices. 
It has been argued that  this scenario leads to inconsistencies even at low energies \cite{peskin}.

The conflict between the Hawking process and quantum mechanics
can be understood in a simple way as follows. 
The ``in'' wave function decomposes
in a part which goes inside the black hole and a part which goes outside the black hole:
\begin{equation}
|\Psi_{\rm in}\rangle \to |\Psi_{\rm BH}\rangle \times |\Psi_{\rm out}\rangle\ .
\label{difo}
\end{equation}
Assuming that the evolution is, as in ordinary  quantum mechanics, linear with a unitary S matrix, after the black hole has disappeared one would have
\begin{equation}
|\Psi_{\rm out}\rangle =S|\Psi_{\rm in}\rangle\ .
\end{equation}
Linearity then implies that 
$|\Psi_{\rm BH}\rangle $ is independent of the initial state, which seems
absurd, unless the wave function is modified after crossing the horizon.
But this would contradict the fact that the horizon is a global construct
and nothing should happen to matter in crossing it.

Note that this contradiction can also be presented in the time $t$ coordinate system 
which is appropriate for the external observer. 
For an evaporating black hole, matter crosses the horizon at finite time $t$.
Time $t$ becomes infinity at a second (``ultimate") horizon, which is a null surface 
inside the black hole and a Planck-order proper distance inside  the event horizon
(the location of the different horizons of an evaporating black hole
is calculated in \cite{hori}). 
In this coordinate system, the decomposition (\ref{difo}) takes place at $t=t_{\rm endpoint}$,
and the above argument applies just in the way it was presented.

A logically possible scenario is one where the black hole does not evaporate completely, so that information remains in a black hole remnant. Hawking radiation could stop when quantum gravity corrections become important. 
However, they become important  when the black hole is Planck size. By this time, most of the energy has been radiated away. 
It does not seem possible that a Planck size object can contain the information of a macroscopic object
(e.g. the information of the sun in $10^{-6}$ gr!). Moreover,
for the information to be preserved, there should be different 
remnants associated with all possible initial states, implying an infinite density of states. Virtual remnants would then affect any quantum process.

\smallskip

An approach proposed by 't Hooft \cite{hooft} consists in  assuming 
that a unitary S matrix exists, and thereby to
deduce what kind of quantum gravity physics this assumption requires.
Unitarity seems to require non-trivial Planck scale processes involving
interactions between ingoing and outgoing modes 
near the horizon which 
would correct the Hawking spectrum. 

If such interactions restore the quantum information encoded in the Hawking radiation, 
then the question is how the process is viewed in an inertial free-falling coordinate system.
In this system there is no Hawking radiation at all, and on semiclassical grounds one does not
expect that matter will be disturbed in crossing the horizon.
This led to Susskind, Thorlacius and Uglum  \cite{susskind} to formulate a principle called Black Hole Complementarity,
which states that information is encoded in the Hawking radiation, but there is no contradiction with
the fact
that observers in free fall are not affected in passing through the horizon of a large black hole.
The degrees of freedom behind the horizon do not commute with those in the Hawking radiation, so 
there is no way one can detect duplication of information (see e.g. \cite{sussk,preskill} for further discussions).

A strong indication that Quantum Gravity must be unitary comes from the AdS/CFT correspondence 
\cite{malda,gubser,witten}. This establishes the equivalence between some conformal field theories, which are known to be unitary,
and Superstring theory on anti de Sitter backgrounds. Since Superstring theory contains gravity, the CFT should describe black holes
within a unitary framework.

The AdS/CFT correspondence, however, does not explain how the paradoxes pointed out above are resolved.
In particular, how the information is retrieved from  black holes, or how the S-matrix can be unitary and
at the same time describe processes involving black hole horizons. It is not either clear how black holes appear in the CFT,
or, more generally, what is the precise way that the CFT encodes gravitational bulk physics.
Some of these issues are addressed by Maldacena in \cite{malda2} and in a recent proposal by Hawking \cite{hawking}. 
The idea is that unitarity in Quantum Gravity is restored after summing over all topologies.
This sums includes a trivial topology contribution, and the claim is that such contribution is relevant (and crucial for unitarity)
even when black hole formation is expected on classical grounds.

\smallskip


A problem which is closely related to the information problem is to understand why the statistical entropy is finite and given by
\begin{equation}
S_{\rm BH}={A_h\over 4G_N}\ .
\end{equation}
In physics, the entropy of a system represents the logarithm of the number of quantum states with given energy.
The question is which are the Quantum Gravity states that give rise to the black hole entropy. Understanding this requires Planck scale physics. 
A naive field theory computation gives an infinite entropy associated with modes near the horizon. In quantum field theory, the horizon acts as an infinite sink that can absorb arbitrary amounts of information.

These contradictions can be avoided by postulating a  
 ``Holographic principle" \cite{hofol}, which states that, 
in Quantum Gravity, the number of orthogonal quantum states inside a given volume is proportional 
to the area of the surface enclosing this volume expressed in Planck units.
The idea is that the black hole entropy provides an upper bound for the entropy of any system that can be put inside
the same surface (see also \cite{susskol}). 
%
%
The statistical derivation of the black hole entropy is a highly non-trivial test for any candidate for a quantum gravitational theory, since it involves counting of physical states in the Hilbert space of Quantum Gravity.

\section{String theory }
\setcounter{equation}{0}

A key ingredient in the AdS/CFT correspondence are the D branes. 
At weak coupling $g_s\ll 1$, they admit a simple description in terms of free open strings with Dirichlet boundary conditions.
At low energies, the open string description reduces to a gauge theory.
The low energy theory of the dynamics of $N$ D branes is described by supersymmetric $U(N)$ Yang-Mills theory.
At strong coupling $g_s\gg 1$, gravitational forces become important and  D branes are expected to become black  branes.
Thus one has two dual alternative description of D branes, in terms of supergravity solutions with closed strings propagating in this supergravity background, or the weak coupling description in terms of Dirichlet strings.

By means of this duality one can provide a microscopic derivation of the black hole entropy in the following way.
For supersymmetric configurations, the degeneracy of the ground state does not depend on the coupling $g_s$, so it can be computed explicitly in the weak coupling limit $g_s\ll 1$, where one has a simple free string theory description and the counting of states is straightforward.
The idea is to consider a supersymmetric  configuration of D branes such that at strong coupling $g_s$ it becomes a black hole with large horizon  radius. Then, the logarithm of the degeneracy of the ground state must give the black hole entropy
with the correct coefficient, since this quantity is independent of the coupling and can therefore be extrapolated to strong coupling where the system represents a black hole.
 
An example is a system containing a large number of  D1 and D5 branes, with momentum along the D1 brane \cite{strova}.
The D1 brane is placed along one of the D5 brane directions. The dual gravity description, after dimensional reduction in the five D5 brane directions,
describes a five-dimensional extremal black hole with three charges $Q_1,\ Q_5,\ N$ (corresponding to the number of
D1  and D5 branes, and momentum, respectively), and finite entropy given by
\begin{equation}
S={A_h\over 4G_N}=2\pi \sqrt{Q_1Q_5N} \ .
\end{equation}
It coincides exactly with the entropy computed by Strominger and Vafa
in \cite{strova} using D brane
technology.
The massless modes are described by a two-dimensional (4,4)
superconformal field theory, governing the low energy dynamics of the
D brane system. The BPS state is a state that contains only left (or
right)
moving excitations, and the degeneracy can be computed as the number
of ways
left moving particles
of $4Q_1 Q_5$ (boson and fermion) species can carry
momentum $N$.
 This highly non-trivial test of string theory has been extended to other D brane configurations.

\medskip

A unitary description of ``black hole evaporation" is in terms of emission of closed strings from near supersymmetric D branes (see e.g. \cite{Kleb,Das}). Note that there is not really a ``black hole evaporation", since in the weak coupling limit where calculations can be done the geometry is flat, in particular, there is no horizon.
The duality provides two different approaches to compute cross sections.
The absorption cross section can be computed either in the $U(N)$ gauge theory, or by solving the field equation in the gravity background. One finds perfect agreement for emission of low energy modes.
The underlying reason for this agreement is due to non-renormalization
theorems that hold for certain correlation functions \cite{mald}.

The duality described above is the basis for  the AdS/CFT correspondence. This general term applies for 
many examples of equivalences between conformal field theories and String theories on anti de Sitter backgrounds.
One of these is the equivalence between $n=4$ supersymmetric $U(N)$  Yang-Mills  theory and Superstring theory on 
$AdS_5 \times S^5$.

As mentioned earlier, the unitary Yang-Mills framework ensures that correlation functions in the Superstring theory
will be unitary, even when they describe strong gravity processes. 
The unitary Yang-Mills framework does not resolve 
the black hole information puzzle for several reasons. One of the 
 problems is that
it is not clear that black holes contribute as intermediate states in scattering amplitudes.
One clue would be to show that the spectrum computed from S-matrix elements approaches
a thermal spectrum in situations where black hole formation is expected. We will return to these points later.

\section{Summing over topologies}
\setcounter{equation}{0}

An interesting idea has been recently proposed by Hawking 
\cite{hawking}, based on previous work by   Maldacena  \cite{malda2}.
Amplitudes are formally defined in Euclidean Quantum Gravity as the path integral over all positive definite metrics that go between two surfaces that are a distance $T$ apart at infinity (Lorentzian amplitudes can then be obtained
by analytic continuation). 
The path integral is taken over metrics of all possible topologies that fit between the surfaces. Typically, one
has a sum of the form
\begin{equation}
A(p_i)
= A_{\rm trivial \ top.}(p_i) +
A_{\rm BH}(p_i)+...
\label{saka}
\end{equation}
Trivial topologies can be foliated by surfaces of constant time. In this case, the time evolution is generated by a Hamiltonian and therefore the evolution should be unitary.
The non-trivial topologies cannot be foliated by surfaces of constant time. For the Euclidean black hole, there is a fixed point at the horizon. Any initial perturbation falls through the horizon, and decays exponentially outside. 
At large times, correlation functions in the black hole geometry will be exponentially small and will not contribute to the scattering amplitude in the sum over topologies  (\ref{saka}). The proposal is that the trivial topology contribution,
even in situations when it is small, provides the necessary contribution to have a unitary S matrix.



\medskip

To make a more precise calculation, one proceeds as follows \cite{hawking}.
To avoid initial and final boundary conditions in the path integral, one identifies the euclidean time: $\tau=\tau+\beta $.
So one is computing the partition function for gravity at finite temperature.
It is divergent in asymptotically flat space, but it is well defined in $AdS_4$. 
Anti de Sitter space acts effectively as a finite box.
Massive particles are confined by the gravitational potential.
Massless particles can get to infinity but they are infinitely redshifted (one can put reflecting walls at infinity which is equivalent to say that there is an identical incoming flux).
$AdS_4$ is the following space-time:
$$
-X_0^2-X_4^2+X_1^2+X_2^2+X_3^2=-R^2\ .
$$
It has boundary $S^1\times S^2$. So one is to sum over all geometries that fit inside this boundary.
One contribution is periodically identified $AdS_4$ with topology 
$S^1\times D^3$   (3-disk). This is a trivial topology, which can be foliated by surfaces of constant time. 
Another contribution is the black hole, with topology $S^2\times D^2$ .

Let us briefly review the thermodynamics of black holes in $AdS_4$ \cite{hawpa}.
$AdS_4$ is described by the following metric:
\begin{equation}
ds^2_{AdS}
= - (1+{r^2\over b^2}) dt^2 + (1+{r^2\over b^2})^{-1}dr^2 +r^2 d\Omega^2 \ ,\ \ \ \ |\Lambda |=3b^{-2}\ .
\end{equation}
The black hole in $AdS_4$ is described by
\begin{equation}
ds^2_{BH}
= - f(r) dt^2 + f^{-1}(r)dr^2 +r^2 d\Omega^2 \ ,\ \ \ \ f(r)=1-{2m\over r}+{r^2\over b^2} \ .
\end{equation}
The horizon is at $r=r_+$\ , with $f(r_+)=0$. The Euclidean solutions are obtained as usual by introducing $\tau=i t$.
The Hawking temperature is given by
\begin{equation}
\beta={1\over T_H}={4\pi b^2r_+\over b^2+3 r_+^2}\ .
\end{equation}
{}For small $r_+$, the temperature is $\sim 1/r_+$. Then, as $r_+$ is increased, the temperature  decreases and it has a minimum
$T=T_0={\sqrt{3}\over 2\pi b}$  at
$r_+=r_0={b\over \sqrt{3}}$. 
In this region $r_+<r_0$ the specific heat is negative and black holes are thermodynamically
unstable. For $r_+>r_0$, the temperature increases monotonically, and the specific heat is positive.
Such ``large" black holes  can be in stable equilibrium with thermal radiation.
This property distinguishes large black holes in $AdS_4$ from black holes in Minkowski space.

The thermodynamics can be summarized as follows:

\smallskip

\noindent $\bullet $ For $T>T_0$, there are two black holes of different masses that contribute to the 
canonical thermal ensemble, but the lower mass black hole is unstable (it decays into radiation or into the other black hole).

\smallskip

\noindent $\bullet $ For $T<T_0$, 
the only contribution to the canonical thermal ensemble is pure radiation, since there is no black hole with that temperature.

\smallskip

\noindent $\bullet $ One can find the action difference from the free energy (AdS is identified with $\tau =\tau + \beta $ )
\begin{equation}
\Delta F=-T\log Z=T\big( I(BH)-I(AdS) \big)\ .
\end{equation}
Using
\begin{equation}
\Delta E=\Delta F+TS={r_+\over 2} (1+{r_+^2\over  b^2})\ ,\ \ \ \ \ S={A_h\over 4}=\pi r_+^2\ ,
\end{equation}
one finds 
\begin{equation}
I(BH)-I(AdS) ={\pi r_+^2(b^2-r_+^2)\over b^2+3 r_+^2}\ .
\end{equation}
So the free energy changes sign at $r_+=b$.
In the region $r_0<r_+<b$, the higher mass black hole is locally stable but it can tunnel to pure radiation, which has
lower free energy, with an amplitude proportional to $\exp(-\Delta I)$.
For $r_+>b$, the black hole configuration is thermodynamically more favorable and black holes dominate the thermal ensemble.

\smallskip

\noindent $\bullet $ There is also a critical energy $E_{\rm collapse}$ such that for $E>E_{\rm collapse}$ only black holes are possible because the thermal ensemble is unstable against gravitational collapse. This critical energy can be seen only by considering self-gravitation effects of the radiation.
Note that one can either work with the canonical ensemble or with the microcanonical ensemble
by integrating over
$d\beta \ e^{\beta E}$ along the imaginary axes (from $-i\infty $ to $+i\infty $). The gravitational instability of
the ensemble corresponds either to some critical energy (microcanonical) or to some critical temperature (canonical).

\medskip

Now suppose we consider some correlation function, 
$$
\langle O_1(\tau_1, x_1)...O_n(\tau_n,x_n)\rangle 
=\langle O_1(\tau_1, x_1)...O_n(\tau_n,x_n)\rangle _{S^1\times D^3}
+\langle O_1(\tau_1, x_1)...O_n(\tau_n,x_n)\rangle _{S^2\times D^2}
+...
$$
In \cite{malda2}, in the context of $AdS_3$, a particular two-point function was computed.
It was found that the correlation function in the black hole background  decays exponentially, 
and that the trivial topology contribution, even if very small, can be consistent with unitarity.
The argument of Hawking generalizing this result is essentially the following one.
For the trivial topology, any global symmetry leads to conserved global charges, which prevent correlation functions from decaying exponentially with time. 
This supports the fact that the evolution on the topologically trivial metrics should be unitary. 
The $AdS_4$ black hole contributes to the second term.
Here there is no conserved quantity, and nothing prevents correlation functions from decaying exponentially. As a result, this piece gives no contribution.
But the total result should be unitary thanks to the trivial topology contribution.

There are some obvious loopholes.
In general, the presence of global charges does not, of course,  guarantee that the S matrix is unitary
(and, moreover, there may be no conserved global charge to start with),
but it is plausible that in the present example of 
the thermal ensemble in $AdS_4$ 
correlation functions will be unitary, as
this is supported by the AdS/CFT correspondence.
One key question is whether in the zero temperature case
correlation functions computed in the topologically trivial
 spacetime will be 
unitary even when black hole formation is expected, i.e. when the
momentum parameters of the initial state are such that collapse is
(classically) inevitable.
In such a situation one would expect that correlation functions  
in the topologically trivial spacetime are exponentially small.
In the finite temperature  case, 
at $r_+>b$, black holes dominate the thermal ensemble and
 correlation functions in the trivial spacetime are exponentially suppressed by a factor 
$ \exp(- \Delta F/T)$ relative to the contribution of the black hole spacetime.
It was argued in \cite{malda2} that even such exponentially small correlation functions can be
consistent with unitarity. The conflict with unitarity can be sharper
when the energy of the system is greater than $E_{\rm collapse}$, i.e. when the energy  is so large that the 
radiation inevitably collapses due to self-gravitation effects. Under such situation, one  expects that correlation
functions on the trivial spacetime are essentially zero, and a paradox remains on how this can be consistent
with unitarity.

Another key question is whether in the problem of black hole formation the Hawking particles arise from the trivial topology contribution, i.e. whether 
the emission spectrum obtained from the S matrix elements (see also section 4.3) 
\begin{equation}
\Gamma (\omega )=\sum_{p_1...p_n}
\langle \Phi_{\rm in}|S^\dagger a^\dagger_\omega |p_1...p_n\rangle _{\rm triv.top.}\ 
\langle p_1...p_n | a_\omega S |\Phi_{\rm in} \rangle _{\rm triv. top.}
\label{werz}
\end{equation}
will approximate a thermal spectrum in a situation when 
gravitational collapse is expected.
If the spectrum is never approximately thermal,
this  probably implies that  black holes are never formed in the theory (at least, not from pure states).
The comparison with a thermal spectrum can  actually  be subtle, as it is illustrated below
by a string theory example.

\section{Alternative scenarios}
\setcounter{equation}{0}

\subsection{Final state boundary condition}

An interesting proposal by Horowitz and Maldacena \cite{hormal} is to impose final state bondary condition at black hole singularities. In this way one avoids the problem of information loss because now one does not need to sum over all possible final states in the interior.
The idea is that
there is a unique wave function at the singularity, similar to Hartle-Hawking unique wave function for the Big-Bang singularity. The wave function must be a superposition of many wave functions representing distinct macroscopic  
states. When matter in a given macroscopic state collapses, the role of the singularity is to precipitate the wave function into the unique wave function. The information is then transferred to the outside by a phenomenon similar to quantum teleportation.

The main problem of this approach is that it relies on unknown physics at the singularity, and 
it is not  known how to construct or  compute this wave function. 
Also, it is not clear that the final state  boundary condition does not enforce the macroscopic state to be of a specific form, leading to a modification of the  semiclassical physics up to the horizon.
 In the paradox of section 1, we mentioned that unitarity along with linearity of the evolution implied that the wave function $|\Psi_{\rm BH}\rangle $ did not depend on the matter that formed the black hole. In a sense, this proposal accepts this.

\subsection{ Do pure states  form black holes ?}
\setcounter{equation}{0}	

The idea that pure states may not form black holes was first proposed by Myers \cite{myers} and by Amati \cite{amati}, 
and more recently by Mathur \cite{mathur}.
In this picture, the emission spectrum in a unitary theory can be approximately thermal only if the initial state is an average,
rather than a pure state.
This is supported by a string-theory calculation in \cite{amatirusso}, showing that the radiation from
a single fundamental string  is exactly thermal if one averages over all initial microscopic states of the same mass (see below).

Can pure states form black holes? Pure states have zero
  entropy. This suggests that
they cannot be associated with a geometry with a horizon.
The idea is that  microstates should not have horizons individually. The notion of horizon should arise only after averaging over the $e^S$  microstates.
Mathur \cite{mathur} proposes that the geometry associated with each microstate is not a black hole. It is similar
outside but it differs significantly in the interior, in particular, these geometries  do not have horizons.  
One example is the
D1-D5-$N$ brane system, although in this case the conjectured geometries for the ``microstates"  are not known.
Amati \cite{amati}, instead, proposes that the notion of the geometry arises only after averaging quantum microstates,
i.e. that it is meaningless to associate a classical geometry with each individual quantum microstate.

The proposal that pure states may not form black holes has some advantages, but it also has some  important weak points: 

\smallskip
\noindent
1. Classically, black holes should be formed even with pure states once a trapped surface forms. There is no explanation why black holes will not be formed.
Indeed, classical gravity does not care whether a given configuration is a pure state or not.
It only sees its energy-momentum tensor.

\smallskip
\noindent 
2. There is so far no clue that the ``coarse graining'' over
   microstate metrics gives (and in what sense) something with a
   horizon. The horizon is a singularity; a ``coarse graining''
over a finite
   number of regular microstates can never give something singular.
In this picture the notion of ``black holes'' would only be an 
approximation to  some regular average.

\smallskip

The main advantage is that the outgoing spectrum is likely to be thermal, almost by construction.
In a unitary theory, the emission spectrum of a pure state is not thermal. If we think of the black hole as a thermal ensemble of the microstates, then a thermal spectrum should arise from the average over the $e^S$ microstates.
 This is true for example in string theory, even if the initial state is a single particle state, as shown in \cite{amatirusso}. 
To see this, one considers the decay rate of a pure massive string state:
\begin{equation}
|\Phi_{M'}\rangle \to |\Phi_{M}\rangle +{\rm graviton}\ .
\end{equation}
The spectrum $\Gamma (\omega )$ is obtained by summing over all final states of given mass $M'$ . In some specific examples, the emission spectrum can be  computed exactly, and it is not thermal.

Now  average over all initial states of mass $M$.
In the case of 
open strings, one finds 
\begin{equation}
d\Gamma (\omega )\cong {\rm const. }
{1\over e^{\omega\over T_{\rm Hag}}-1}\omega^{d-2} d\omega \ ,
\end{equation}
{}which shows that open strings behave as a perfect black body at the Hagedorn temperature $T_{\rm Hag}$
(see also \cite{cir2} for the superstring calculation).
For closed strings, one finds 
\begin{equation}
d\Gamma (\omega ) \cong {\rm const.} M 
{1\over e^{\omega\over T_{L}}-1}{1\over e^{\omega\over T_{R}}-1} \omega^{d-1} d\omega ={\rm const.} M
{\sigma (\omega ) \over e^{\omega\over T}-1}\omega^{d-2} d\omega \ ,
\end{equation}
where 
$$
T^{-1}=T_L^{-1}+T_R^{-1}\ , \ \ \ \ T_{L,R}={2\over M }\sqrt{M^2-Q_{L,R}^2} \ ,
$$
and $ \sigma (\omega ) $  is the grey body factor
\begin{equation}
\sigma (\omega ) ={\omega\ (e^{{\omega\over T}}-1)\over (e^{{\omega\over T_L} }-1)(e^{{\omega\over T_R}}-1)}\ .
\end{equation}
How does the spectrum look like if we do not average over initial states?
For a generic initial state, the spectrum is very far from thermal, in particular, in general it does not  have an exponential tail. However, there is a special string state whose spectrum resembles a thermal spectrum, 
$$
|\Phi \rangle = (b^\dagger_{1R})^N (c^\dagger_{1L})^N |0\rangle\ ,
$$
where $b_n$ and $c_n$ are the mode operators in the planes
$X_1+iX_2$ and $X_3+i X_4$.  This quantum state was investigated in \cite{cir2}
because it is remarkably stable. It describes a `` rotating ring", 
the corresponding classical configuration is a rigid, circular string with angular momentum in two orthogonal planes.
Now consider the spectrum for massless emission
\begin{equation}
\Gamma (\omega )=\sum_{f\in \hat N=N'} \langle \Phi_{\rm in} | 
V^\dagger (\omega )|\Phi_f\rangle
\langle \Phi_f| V(\omega )|\Phi_{\rm in}\rangle \ .
\end{equation}
This calculation was carried out in \cite{cir2}.
The resulting spectrum is not thermal, as expected. 
But it has a maximum and an exponential tail $\Gamma(\omega )\sim
\exp(-\omega /T)$ with $T={a_0\over \alpha' M},\ a_0^{-1}=2\log 2-1$.
The ``thermal"  shape
seems  to be special to this state, and it is related to the fact that the dominant channel of decay is by radiation
(for large masses, the spectrum can be reproduced by modelling the ring as a classical radiating antenna).

\subsection{Separating ``thermal" part in S-matrix contributions}

Let us now comment on a different strategy for a possible resolution of the information puzzle.
It has some similarities with  section 3, in the sense that 
one needs to add all Feynman diagrams of different topology in order to restore unitarity, but only
a subset of the Feynman diagrams represent emission from black holes.
The thermalization of the spectrum is produced by a mechanism similar to that of \cite{amatirusso}, but
instead of introducing the average over microstates by hand, we claim that this average is effectively produced in summing
over soft radiation modes emitted just before black hole formation.

Consider an ``in" state describing a set of  incoming particles. 
For simplicity, we will assume that it is composed of two particles.
We consider the case where the center of mass energy is much larger than the impact parameter,
so that on classical grounds the system will form a black hole (see for example \cite{giddings,ven}).
%
%
The spectrum computed from the $S$ matrix is given by
\begin{equation}
\Gamma (\omega )=\sum_{p_1...p_n}
\langle \Phi_{\rm in}|S^\dagger a^\dagger_\omega |p_1...p_n\rangle 
\langle p_1...p_n | a_\omega S |\Phi_{\rm in} \rangle \ .
\label{wezzg}
\end{equation}
In a consistent quantum theory,  
the total sum of Feynman diagrams must give a unitary result, i.e.
\begin{equation}
\sum_{p_1...p_n}
\langle \Phi_{\rm in}|S^\dagger|p_1...p_n\rangle 
\langle p_1...p_n |  S |\Phi_{\rm in} \rangle =
\langle \Phi_{\rm in}|S^\dagger   S |\Phi_{\rm in} \rangle = 1\ .
\label{unita}
\end{equation}
Under these conditions, we conjecture  that the spectrum $\Gamma(\omega ) $ given in
 (\ref{wezzg}) is not thermal (not even in an approximate sense) in any region of the phase space.

{}To justify this conjecture, consider figures 1 and 2. The state $|\Phi_{\rm in}\rangle $ 
is composed of two particles $\Psi $ and $\xi $.
For concreteness, one can imagine that the state $\Psi $ is a very massive state
which will become a black hole after receiving the energy carried by
the wave $\xi $.
The spectrum (\ref{wezzg}) contains the contribution of Feynman diagrams
such as fig. 1, where the radiation of the mode $\gamma $  occurs
before black hole formation. 
In the Feynman diagram of fig. 2, the 
 mode $\gamma $ is emitted after $\Psi $ receives the energy of the state $\xi $, i.e. from a state $\Sigma $ formed by the collision of $\Psi$ and $\xi $. 
Since by assumption the state $\Sigma $ has a size less than its gravitational radius, one can expect that this state 
will behave like a  black hole.
These contributions  should give a spectrum which is very close to a thermal spectrum, that is
\begin{equation}
\Gamma (\omega )=\sum_{p_1...p_n}
\langle \Phi_{\rm in}|S^\dagger a^\dagger_\omega |p_1...p_n\rangle _2
\ \langle p_1...p_n | a_\omega S |\Phi_{\rm in} \rangle _2\cong  {\sigma (\omega )\over 
e^{\omega\over T_H}-1}\omega^{d-2} d\omega\ .
\label{wezh}
\end{equation}
The subindex ``2" indicates the contractions of field operators giving
rise to diagrams like fig. 2. More precisely, diagrams where
the state $\Sigma $ that emits $\gamma $ has an energy greater than
the impact parameter (in Planck units).
The origin of the thermalization would be
the sum in (\ref{wezh}) over Feynman diagrams describing emission prior to the emission of the mode $\gamma $.
It is possible that the net effect of these  sums is to replace the state
$\Sigma $ by an average over a set of quantum states 
with the same energy and a size  less or equal than the impact parameter.
The reason is that the sum in  (\ref{wezh}) contains all possible
emission  processes and the contributions of
many different  quantum states $\Sigma $ of a given energy.
The average over quantum states should then produce a thermal spectrum 
as in \cite{amatirusso}.

Note that some transitions from $\Psi $ to some  quantum
states might be very suppressed, so in general the sums 
will not produce an average with equal weights.
This is precisely the mechanism that should select a specific set of 
quantum microstates contributing to the average.
Amplitudes for the initial quantum state to make transitions into 
this set of quantum microstates 
by emission of soft radiation should be of similar order.
Then this would produce the effective average state $\Sigma $ that emits thermal radiation. 


\begin{figure}
\center
\includegraphics[height=5cm]{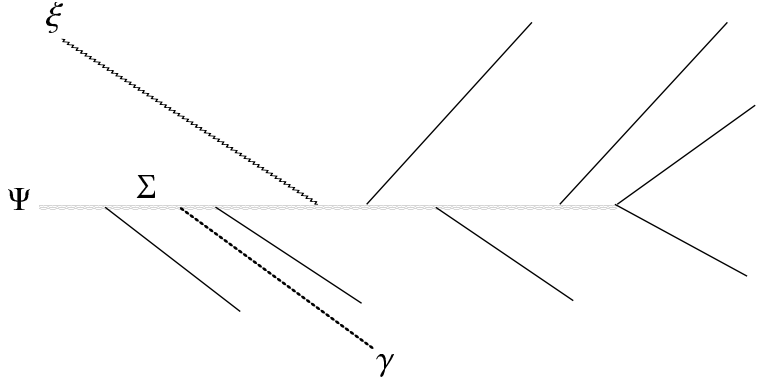}
\caption{The mode $\gamma $ is emitted before $\Psi $ absorbs $\xi $.
The solid lines represent soft outgoing radiation.}
\label{figure1a}       
\end{figure}

\begin{figure}
\center
\includegraphics[height=5cm]{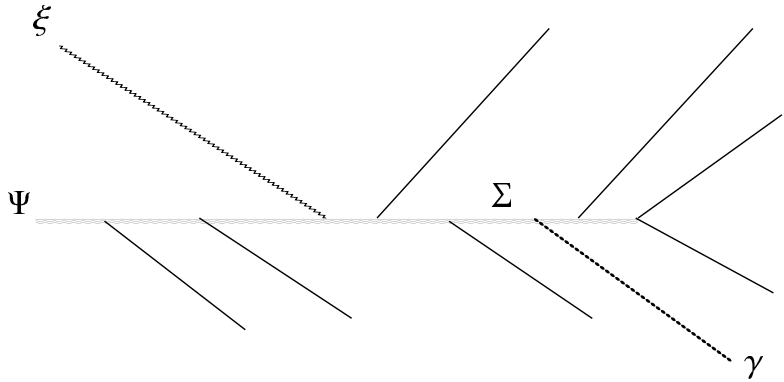}
\caption{The mode $\gamma $ is emitted after $\Psi $ absorbs $\xi $.}
\label{figure1b}       
\end{figure}

\begin{figure}
\center
\includegraphics[height=5cm]{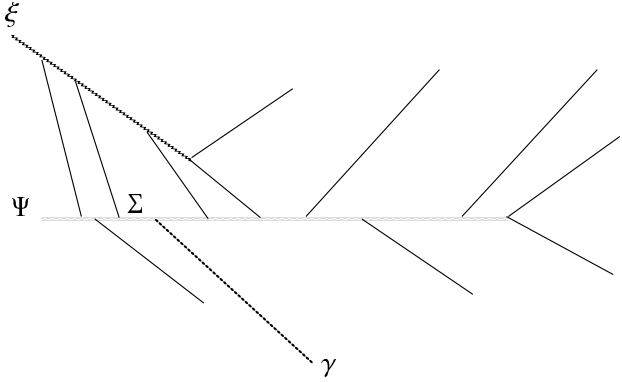}
\caption{A loop correction to the amplitude.}
\label{figure2}       
\end{figure}

The thermal
spectrum is expected to arise only when strong gravitational effects
have been taken into account,
when all relevant loop contributions have been resummed.
To loop level, the contributions of diagrams describing $\gamma $
emitted ``before" black
hole formation cannot be isolated just from the topology of the
Feynman  diagram.
From the viewpoint of tree-level Feynman diagrams, there is a simple
distinction between diagrams 1 and 2, i.e. emission  taking place 
before or after  
 $\Psi $ absorbs the energy of $\xi $.
But once loop corrections are incorporated, the momentum transfer takes place in several steps.
An example of loop corrections is shown in fig. 3.
Since one integrates over the momenta of the loops, the same Feynman
diagram contains contributions where $\Sigma $ carries small or large
amounts of energy. 

Instead of trying to learn something from the full S-matrix describing
black hole formation and evaporation, which is something extremely 
complicated, let us see what can be learned from the semiclassical
picture.
First,  the issue of emission  ``before" or ``after" black hole formation 
needs some explanation.
Classically, to an external observer,  infalling collapsing matter
never crosses the horizon, so there is not a sharp time where she can state that the black hole has been formed. 
Semiclassically, matter crosses the horizon at some given (late) time. 
To the external observer, the black hole appears to be formed when she
gets into causal contact
with the apparent horizon (which for an evaporating black hole
has a part which lies outside the even horizon \cite{hori,modelbh}).
It is by this time that the process of Hawking radiation begins to be important.
Indeed, the total energy radiated before this time was calculated in
\cite{modelbh}
and gives $E_{\rm out}^{\rm bef}=O\big( (mG)^{-1}\big)$.
%
%
The issue is whether this small amount of radiation
is enough to thermalize the initial state. 
Since Hawking modes carry an average energy  $\sim T_H\cong (mG)^{-1}$,
this suggests that only a few Hawking particles are emitted
before the observer gets into causal contact with the apparent horizon.
However, as pointed out above, one has also to sum over the emission of all
soft modes with energies $<(mG)^{-1}$. It is possible that this  produces the transitions between the initial
state and all possible microstates of the black hole, thermalizing
the quantum state before the great volume of Hawking emission starts.

Some aspects of this approach can be tested by simple calculations in Superstring theory.
 One can consider as the ``in" state a massive superstring state $\Psi $ and a graviton $\xi $ carrying some energy.
Then one considers the emission of one or more massless modes and compute the spectrum (\ref{wezh}) obtained
from fig. 2
(averages of Feynman diagrams describing one absorption and one emission of a massless mode are computed in \cite{manes}
in a study of string form factors).
The idea of this calculation would be to see if the sum over Feynman diagrams and phase space involved in (\ref{wezh})
can be effectively  replaced by an average over microstates and whether the resulting emission spectrum is close to
a thermal spectrum.

\section{Conclusions}

The main points that we have discussed in this review are the following ones:

\smallskip
\noindent $\bullet $ The AdS/CFT duality is a strong evidence that a unitary S matrix exists in quantum gravity.
But so far there is not a good understanding on how the CFT
encodes the bulk gravitational physics.

\smallskip
\noindent $\bullet $ It would be important to show if and how black holes contribute as 
intermediate states in any correlation function, and to investigate if Hawking radiation can be derived from correlation functions.

\smallskip
\noindent $\bullet $ A strategy 
to solve the information paradox is to separate
the  contributions to the S-matrix that represent emission from a black hole. 
This part would give a spectrum which is close to a thermal spectrum.
Unitarity of the S matrix would be seen only after adding all contributions.


\section*{Acknowledgements}

I would like to thank D. Amati and  G. Veneziano for discussions 
and P. Townsend for a remark.
This work is partially supported by MCYT FPA,
2001-3598 and CIRIT GC 2001SGR-00065, and
by the European Comission RTN program under contract
MRTN-CT-2004-005104.

%
%
%

\end{document}